\documentclass[useAMS]{stylefile}

\usepackage{amsmath}
\usepackage{latexsym}
\usepackage{amssymb}
\usepackage{bm}
\usepackage{graphicx}
\usepackage{textcomp}
\usepackage{natbib}

\def\p{\partial}
\def\b{\textbf}

\def\n{\nabla}
\def\x{\times}

\def\h{\hat}
\def\'{\prime}
\def\<{\left\langle}
\def\>{\right\rangle}
\def\~{\tilde}
\def\ep{\varepsilon}

\begin{document}
\doi{10.1080/03091920xxxxxxxxx}
 \issn{1029-0419} \issnp{0309-1929} \jvol{00} \jnum{00} \jyear{2010} \jmonth{October}

\markboth{Richardson and Proctor}{Effects of fluctuations on nonlinear dynamo models}

\title{\textit{Effects of $\alpha$-effect fluctuations on simple nonlinear dynamo models}}

\author{K. J. RICHARDSON${\dag}$ $^{\ast}$\thanks{$^\ast$Corresponding author. Email: kjr42@cam.ac.uk
\vspace{6pt}} and M. R. E. PROCTOR${\dag}$\\\vspace{6pt}  ${\dag}$ DAMTP, Centre for Mathematical Sciences, University of Cambridge, Wilberforce Road, Cambridge, CB3 0WA\\ \vspace{6pt}\received{Accepted 3rd September 2010} }

\maketitle

\begin{abstract}
We investigate the interaction of a fluctuating $\alpha$-effect with large-scale shear in a simple nonlinear 1-dimensional  dynamo wave model.  We firstly extend the calculations of Proctor [Effects of fluctuation on $\alpha \Omega$ dynamo models. \textit{MNRAS} 2007, \b{41}, L39-L42] to include spatial variation of the fluctuations, and find that there can be a mechanism for magnetic field generation, even when the mean $\alpha$ is zero, provided the spatiotemporal spectrum of the fluctuations has an appropriate form.  We investigate mean-field dynamo action when the new term arising from the fluctuations is non-zero, and present results concerning the stability and frequency of the solutions and parity selection in the nonlinear regime.  The relation between the asymptotic theory and explicit simulation of a traditional mean-field model with a fluctuating function for the $\alpha$-effect term is discussed.

\bigskip

\end{abstract}

\section{Introduction} \label{sec:introduction}

It is widely accepted that the large-scale magnetic field generation in the Sun is governed by a dynamo process.  Mathematical models of this dynamo process standardly decompose the magnetic field into poloidal and toroidal parts, where the aim is to complete the dynamo loop of generating toroidal field from poloidal field and vice versa.  The presence of differential rotation within the Sun's convection zone is believed to be responsible for turning poloidal field into toroidal field, this procedure is known as the $\Omega$-effect.  The mechanism by which poloidal field is generated from toroidal field is still open to debate, and a number of different models have been investigated. One such model is that of `mean-field electrodynamics', where the magnetic  field is supposed to exist on two very different scales, and large-scale magnetic fields are generated through a mean emf induced by the averaged properties  of small scale helical motions (the `$\alpha$-effect') (see \cite{Moffatt78} and \cite{KrauseRadler80} for details).   \\

The $\alpha$-effect is associated with an average over the small scales of the cross-product of the small scale velocity and magnetic fields.  Since it is due to an average, it is generally treated as a non-zero variable varying only on long spatial and temporal scales. This assumption would be  acceptable provided that any high-frequency fluctuations of  $\alpha$  around this mean value are not too large.  A turbulent, helical flow, which lacks reflectional symmetry about the equator, might be thought to provide perfect conditions for the $\alpha$-effect to work, enhancing the growth of large-scale poloidal magnetic field and leading to a large-scale dynamo.  A study by \cite{CattaneoHughes06}, however,  found a contradictory result.  They studied rotating Boussinesq convection where the flow was turbulent and helical, and found no large-scale dynamo.  They calculated the emf and plotted $\alpha$ as a function of time but found it to be wildly fluctuating with a very small long-time average.  \\

Previous work by \cite{Silant'ev00} and \cite{Proctor07} has shown that when such fluctuations in the $\alpha$-effect interact with a large-scale shear, unlike \cite{CattaneoHughes06}, then large-scale dynamo action can occur even when the mean $\alpha$ is zero.  Both of these investigations used an $\alpha$-$\Omega$ dynamo model; \cite{Silant'ev00} considered spatial variation of the fluctuations in quite a complicated calculation, whereas \cite{Proctor07} considered a temporally fluctuating $\alpha$-effect in a much simpler setup.  The \cite{Proctor07} calculation used a one-dimensional dynamo wave model and the $\alpha$-effect was split into mean and fluctuating parts, where the fluctuations were large.  A suitable average was taken, which resulted in a new term involving the fluctuating part of $\alpha$ in the mean field evolution equations.  It was then found that this new term can lead to a large-scale dynamo mechanism, even when the mean $\alpha$ term is zero.  This paper also found that when this new term was included in a simple mean field model of the solar cycle, the effect was a lengthening of the cycle period as the amplitude of the new term increased;  at a sufficiently large value, the solutions became steady.  \\

\cite{HughesProctor09} revisited the calculation of \cite{CattaneoHughes06} with the addition of large-scale shear, and found vigorous large-scale dynamo action.  They concluded that the effect of introducing large-scale shear could be due to one of two mechanisms: the anisotropy of the shear could lead to a shear-current effect as discussed in \cite{RogachevskiiKleeorin08}, or the shear could interact with a temporally fluctuating $\alpha$-effect as in \cite{Proctor07} both of which can be associated with off diagonal components of the turbulent diffusivity tensor, and lead in simple geometries to the same addition term in the mean field equations.  Either of these effects could also explain the amplification of the large-scale magnetic field found in previous numerical studies by \cite{YousefHeinemannetal08} and \cite{KapylaKorpiBrandenburg08}, where 3D simulations of turbulent convection and shear were investigated.  These calculations support the necessity of large-scale shear, and correspond to the case in \cite{Proctor07} where a large-scale dynamo mechanism was found with no mean $\alpha$-effect.\\

The purpose of this paper is to extend the investigation of a fluctuating $\alpha$-effect in \cite{Proctor07} by looking at the linear and nonlinear consequences of the new term in a finite geometry.  Firstly, in section \ref{sec:extension_of_model}, it is shown that the model of \cite{Proctor07} can be extended to include more realistic spatial and temporal dependent fluctuations.  Detailed studies of the dynamics of the model in the linear and nonlinear regimes are conducted in section \ref{sec:numerical_model}, which allow us to understand precisely how the steady and oscillatory modes interact, and the effects of parity on the system.  Preliminary results of this work are discussed briefly in \cite{ProctorRichardsonBushby09}.  Finally, in section \ref{sec:test_with_random_function}, the fluctuating $\alpha$-effect theory is verified by relating it to a simple one-dimensional model with a fluctuating function for the $\alpha$-effect term. \\

\section{Formulation of the model} \label{sec:extension_of_model}

A description of the mechanism in spherical polar coordinates can be found in \cite{Proctor07}. In this paper we use a simplified one-dimensional cartesian version of the model.  We begin with  a one-dimensional Parker dynamo wave model as adapted by \cite{ProctorSpiegel91}.  In this cartesian set-up, the magnetic field is assumed axisymmetric such that $\b{B} = B(x,t) \b{e}_{y} + \n \x [A(x,t) \b{e}_{y}]$, where $x$, $y$ represent the North-South and azimuthal directions respectively, and $A$ and $B$ are the respective poloidal and toroidal magnetic fields.  The governing equations take the form of an $\alpha \Omega$ dynamo and can be written
\begin{subequations}
\begin{eqnarray}
 A_{t} &=& \alpha B + \eta \left( A_{xx}-\ell^{2} A \right), \label{eq:1d_model_A} \\
 B_{t} &=& \Omega^{\prime} A_{x} + \eta \left( B_{xx}-\ell^{2} B \right), \label{eq:1d_model_B}
\end{eqnarray}
\end{subequations}
where $\eta$ is the magnetic diffusivity, $\ell$ is an inverse lengthscale, and subscripts $t$, $x$ denote differentiation with respect to time and space respectively.  In equation \ref{eq:1d_model_B}, $\Omega^{\prime}$ is the differential rotation (large-scale shear) which generates toroidal field from poloidal field, and the term involving $\alpha$ in equation \ref{eq:1d_model_A} is the traditional $\alpha$-effect term, which closes the dynamo loop by generating poloidal field from toroidal field by the action of small-scale helical motions.\\

Spatial and temporal fluctuations are introduced in the $\alpha$-effect term by writing $\alpha$ as a sum of its mean and fluctuating parts, where the fluctuating temporal and spatial scales ($\tau$ and $\xi$ respectively) are longer than that of the original averaging process, but shorter than that of mean-field evolution:
\begin{equation}
 \alpha = \alpha_{0} + \delta^{-2} \alpha_{1} \left( \tau, \xi \right), \label{eq:alpha_expression}
 \end{equation}
where subscripts $0$ and $1$ represent the mean and fluctuating parts respectively, and dependence on $x$ and $t$ has been suppressed.  To ensure the fluctuations are large, $\delta^{2}$ is taken to be small, and corresponds to $\ep$ in the \cite{Proctor07} case where spatial fluctuations are not included.\\

A method of multiple scales is applied such that
\begin{align*}
\p_{t} &\rightarrow \p_{t} + \delta^{-2} \p_{\tau},\\
\p_{x} &\rightarrow \p_{x} + \delta^{-1} \p_{\xi},
\end{align*}
 which induces the following corrections to $A$ and $B$ 
\begin{subequations}
\begin{eqnarray}
A &=& A_{0} + A_{1}(\tau,\xi) + \delta A_{2}(\tau,\xi) + \cdots, \label{eq:A_expansion} \\
B &=& B_{0} + \delta B_{1}(\tau,\xi) + \delta^{2} B_{2}(\tau,\xi) + \cdots. \label{eq:B_expansion}
\end{eqnarray}
\end{subequations}

These scalings are applied to equations (\ref{eq:1d_model_A}, \ref{eq:1d_model_B}), and an average is taken over the intermediate scales, determined by $\< \cdot \>$ such that  $\<\alpha_{1}\> = \< A_{1}\> = \<A_{2}\> = \<B_{1}\> = \<B_{2}\> = 0$, to obtain a set of equations describing the evolution of the mean fields $A_{0}$ and $B_{0}$:
\begin{subequations}
\begin{eqnarray}
{A_{0}}_{t} &=& \alpha_{0} B_{0} + \delta^{-1} \<\alpha_{1} B_{1} \> + \< \alpha_{1} B_{2} \> + \eta \left( {A_{0}}_{xx} - \ell^{2} A_{0} \right), \label{eq:1st_mean_eqns_A} \\
{B_{0}}_{t} &=& \Omega^{\'} {A_{0}}_{x} + \eta \left( {B_{0}}_{xx} - \ell^{2} B_{0} \right), \label{eq:1st_mean_eqns_B}
\end{eqnarray}
\end{subequations}
where $\delta^{-1} \< \alpha_{1} B_{1}\>$ and $\<\alpha_{1}B_{2}\>$ are new terms that arise purely from the fluctuations.  The term $\delta^{-1} \<\alpha_{1}B_{1}\>$ appears to violate the scaling, but vanishes under plausible symmetry assumptions. The mean field equations in this case are given by
\begin{subequations}
\begin{eqnarray}
{A_{0}}_{t} &=& \alpha_{0} B_{0} + \< \alpha_{1} B_{2} \> + \eta \left( {A_{0}}_{xx} - \ell^{2} A_{0} \right), \label{eq:mean_eqns_A} \\
{B_{0}}_{t} &=& \Omega^{\'} {A_{0}}_{x} + \eta \left( {B_{0}}_{xx} - \ell^{2} B_{0} \right), \label{eq:mean_eqns_B}
\end{eqnarray}
\end{subequations}
where $\<\alpha_{1}B_{2}\>$ is the term we wish to calculate to compare with the corresponding term in \cite{Proctor07}, i.e. the term proportional to ${B_{0}}_{x}$.\\

In order to calculate $\<\alpha_{1}B_{2}\>$ we subtract (\ref{eq:mean_eqns_A}, \ref{eq:mean_eqns_B}) from the full equations (\ref{eq:1d_model_A}, \ref{eq:1d_model_B}) (before the averaging) and take leading order terms to find equations that describe the evolution of the fluctuating fields $A_{1}$ and $B_{1}$:
\begin{subequations}
\begin{eqnarray}
{A_{1}}_{\tau} &=& \alpha_{1} B_{0} + \eta {A_{1}}_{\xi\xi}, \label{eq:fluctuating_eqns_A_1} \\
{B_{1}}_{\tau} &=& \Omega^{\'} {A_{1}}_{\xi} + \eta {B_{1}}_{\xi\xi}, \label{eq:fluctuating_eqns_B_1}
\end{eqnarray}
\end{subequations}
and we take the next order terms to find equations that describe the evolution of the fluctuating fields $A_{2}$ and $B_{2}$:
\begin{subequations}
\begin{eqnarray}
{A_{2}}_{\tau} &=& \alpha_{1} B_{1} + 2 \eta {A_{1}}_{x \xi} + \eta {A_{2}}_{\xi\xi}, \label{eq:fluctuating_eqns_A_2} \\
{B_{2}}_{\tau} &=& \Omega^{\'} {A_{1}}_{x} + \Omega^{\'} {A_{2}}_{\xi} + 2 \eta {B_{1}}_{x \xi} + \eta {B_{2}}_{\xi\xi}. \label{eq:fluctuating_eqns_B_2}
\end{eqnarray}
\end{subequations}

Equations (\ref{eq:fluctuating_eqns_A_1}, \ref{eq:fluctuating_eqns_B_1}, \ref{eq:fluctuating_eqns_A_2}, \ref{eq:fluctuating_eqns_B_2}) are solved to find an expression for $B_{2}$, which enables us to calculate $\< \alpha_{1} B_{2} \>$.  This average is of the form
 \begin{equation}\label{eq:alpha_1_B_2}
 \< \alpha_{1} B_{2} \> = \alpha_{f} B_{0} - G \Omega' {B_{0}}_{x} ,
 \end{equation}
where $\alpha_{f}$ is a contribution to the traditional $\alpha$-effect term, and 
\begin{equation}\label{eq:G_expression}
G =  \int \int \frac{\vert \h{\alpha}_{1} \vert ^{2} (\omega^{4} - 12 \eta^{2} \omega^{2} k^{4} + 3 \eta^{4} k^{8})}{(\omega^{2} + \eta^{2} k^{4})^{3}} \mathrm{d}k \mathrm{d}\omega ,
\end{equation}
where $\h{\alpha}_{1} = \h{\alpha}_{1}(k,\omega)$ is the Fourier transform in space and time of $\alpha_{1}$.  There are two contributions to $\alpha_{f}$; one from the expansion, which vanishes under plausible symmetry assumptions, and the other from finite boundary effects, which are uncalculable due to the spatial inhomogeneity of the background field.  We therefore regard $\alpha_{f}$ as a renormalisation of the $\alpha$-effect, and concentrate on the effect of the term $\propto {B_{0}}_{x}$.  Note that the notation is different from \cite{Proctor07}, where the expression corresponding to equation (\ref{eq:G_expression}) (with $k=0$) is positive definite and is denoted by $G^{2}$.  The new feature of equation (\ref{eq:G_expression}) is that when spatial fluctuations are included the expression may take either sign in contrast to \cite{Proctor07}.  More interestingly, we see that $G$ is positive at both small and large $\eta$, and there exists a range for which $G$ is negative; $(2+\sqrt{11/3})^{-1} < \omega^{2}/\eta^{2} k^{4} < (2-\sqrt{11/3})^{-1}$.  Note that the corresponding formula given in \cite{ProctorRichardsonBushby09} is not correct, although the nonlinear calculations are not effected by this.\\

By substituting the expression for $\<\alpha_{1} B_{2}\>$ into equation (\ref{eq:mean_eqns_A}), and dropping the zero subscripts on $A$ and $B$, we have the final version of the model:
\begin{subequations}
\begin{eqnarray}
 A_{t} &=& \alpha_{0} B - G \Omega' B_{x} + \eta \left(A_{xx} - \ell^{2} A \right), \label{eq:A_eqn} \\
 B_{t} &=& \Omega^{\prime} A_{x} + \eta \left( B_{xx} - \ell^{2} B \right), \label{eq:B_eqn}
\end{eqnarray}
\end{subequations}
where we have subsumed $\alpha_{f}$ into the definition of $\alpha_{0}$.  It is interesting to investigate how this new term $G$ effects dynamo action, and whether it alone can lead to large-scale dynamo action with no mean $\alpha$. As indicated above, other possible effects can lead to a term of this nature, such as the shear-current effect, see \cite{RogachevskiiKleeorin08}.\\

The model is solved analytically by seeking solutions $A, B \propto \exp \left[ i \left( k x + \omega t \right) \right]$ to obtain the following dimensionless dispersion relation
\begin{equation}\label{eq:dimensionless_disp_rel}
 \left(m^{2} + 1 \right)^{2} -Q m^{2} = \frac{D^{2} m^{2}}{4 \left( m^{2} + 1 \right)^{2}},
\end{equation}
where $k$ is scaled such that $k = \ell m$, $m$ constant, $Q = G {\Omega^{\prime}}^{2} / (\eta^{2} \ell^{2})$ is the dimensionless form of $G$ and $D = \Omega^{\prime} \alpha_{0} / (\eta^{2} \ell^{3})$ is the dynamo number.  The corresponding dispersion relation found in \cite{Proctor07} has the same form but with different notation, since we now use $Q$ rather than $Q^{2}$ to indicate that this term may now take either sign.\\

\begin{figure}
\begin{center}
\includegraphics[width=7cm]{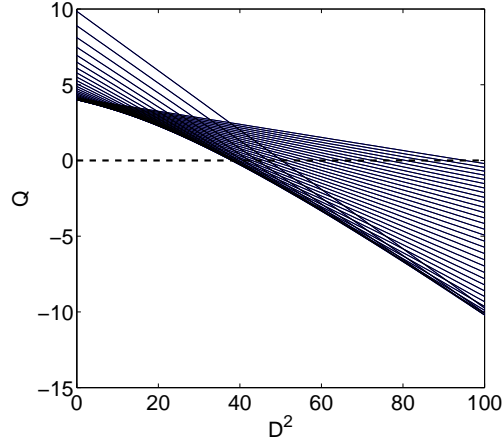}
\caption{Marginal stability boundary, found by plotting lines of constant $m$ from $m=0.36$ to $m=1.2$ for $Q$ against $D^{2}$ according to equation (\ref{eq:dimensionless_disp_rel}).  Dynamo action is possible above the lower envelope of lines, and the region below the thick dashed line at $Q=0$ is the additional region of possible dynamo action found by considering spatial dependence in the fluctuations of $\alpha$.}\label{fig:margin}
\end{center}
\end{figure}

The marginal stability boundary is plotted in figure \ref{fig:margin} by plotting lines of constant $m$ according to equation (\ref{eq:dimensionless_disp_rel}).  Dynamo action is possible above the lower envelope of lines.  This figure has the same form as in \cite{Proctor07} when $Q$ is positive, but due to the spatial fluctuations, $Q$ may be negative and the diagram can be extended as shown.  It can be seen that dynamo action in the negative $Q$ regime is inhibited by the new effect and larger dynamo numbers $D$ are required for instability.  Recall that there is a range for which $Q$ is negative, given above, while $Q$ is positive when $\eta$ is both negligible and when it dominates. Figure 1 tells us that we can achieve dynamo action when the dynamo number is zero, and therefore when the mean $\alpha$ is zero.  In the case when $D=0$, we find that $\omega=0$, and hence dynamo waves do not travel, so that solutions are steady in this region.  To understand the stability of the system, and discover the range over which steady solutions can be found in a finite geometry, we introduce a nonlinear quenching term and solve the equations numerically. 

\section{Numerical Model of the Solar Cycle} \label{sec:numerical_model}

To enable us to solve the model numerically we use the same method as \cite{Proctor07} to simulate a simple model of the solar cycle, and investigate further the stability and parity of the system. We introduce a nonlinear quenching so that the $\alpha$-effect takes the form $\alpha (x,t)/(1+B^2)$. Equations (\ref{eq:A_eqn}, \ref{eq:B_eqn}) are rescaled so that $\Omega^{\prime} = \eta = \ell = 1$.  We also choose $\alpha_{0} = -d \sin(2 \pi x/l)$ to simulate antisymmetric dynamo waves that travel towards the equator, where $d$ is a positive constant, and we parametrise the new term due to the fluctuations with the positive constant $r$.  We are then able to vary the parameters $d$ and $r$ to investigate the stability in $d-r$ phase space both in the linear and nonlinear regimes. The model equations then become
\begin{subequations}
\begin{eqnarray}
A_{t} &=& \frac{- d \sin \left(2 \pi x / l \right) B - r B_{x} }{1 + B^{2}} + A_{xx} -A, \label{eq:numerical_eqn_A}\\
B_{t} &=& A_{x} + B_{xx} -B, \label{eq:numerical_eqn_B}
\end{eqnarray}
\end{subequations}
which are to be solved between $0<x<l$, with $A=B=0$ at $x=0,l$.

\subsection{Linear Model} \label{sec:linear_model}

The linear stability boundary can be found by setting the denominator of the terms in $B$ on the right-hand-side of (\ref{eq:numerical_eqn_A}) to unity, leading to the system 
\begin{subequations}
\begin{eqnarray}
A_{t} &=& - d \sin \left(2 \pi x / l \right) B - r B_{x} + A_{xx} -A, \label{eq:linear_eqn_A}\\
B_{t} &=& A_{x} + B_{xx} -B, \label{eq:linear_eqn_B}
\end{eqnarray}
\end{subequations}
and then performing a linear stability analysis on these equations.  This was done by expanding $A$ and $B$ as Fourier sine series in $x$ and reducing the system to a matrix problem.  This matrix was then decomposed into two individual matrices; one for dipole solutions (entries corresponding to $A$ even and $B$ odd), and one for quadrupole solutions (entries corresponding to $A$ odd and $B$ even.) The complex growth rate for each mode was then determined as an eigenvalue of the appropriate matrix, and the parameters $d$ and $r$ were then varied to find the stability boundaries for each parity, and for each mode of instability. \\

\begin{figure}
\begin{center}
\subfigure[$l=4$]{
\resizebox*{5cm}{!}{\includegraphics{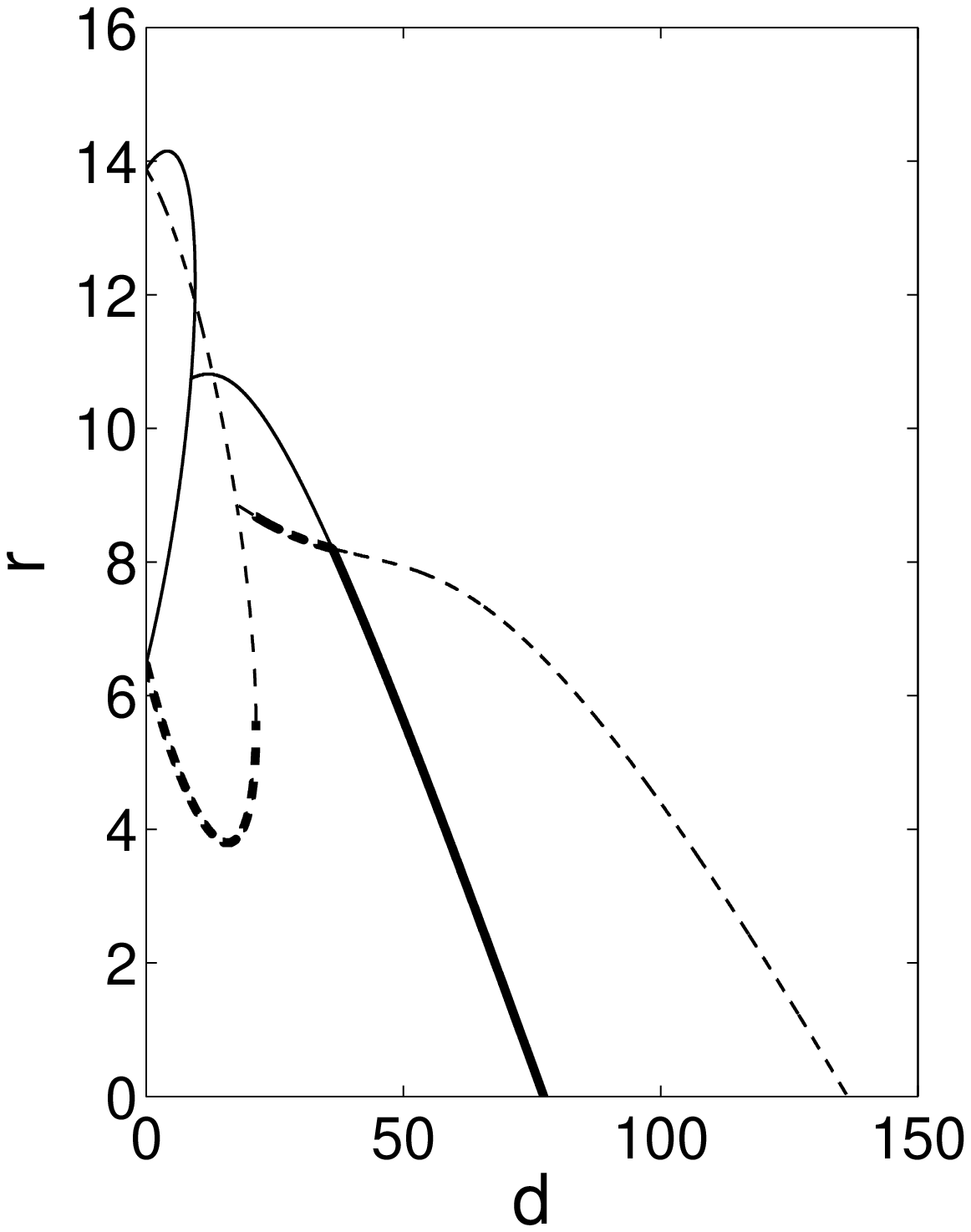}}}%
\subfigure[$l=7$]{
\resizebox*{5cm}{!}{\includegraphics{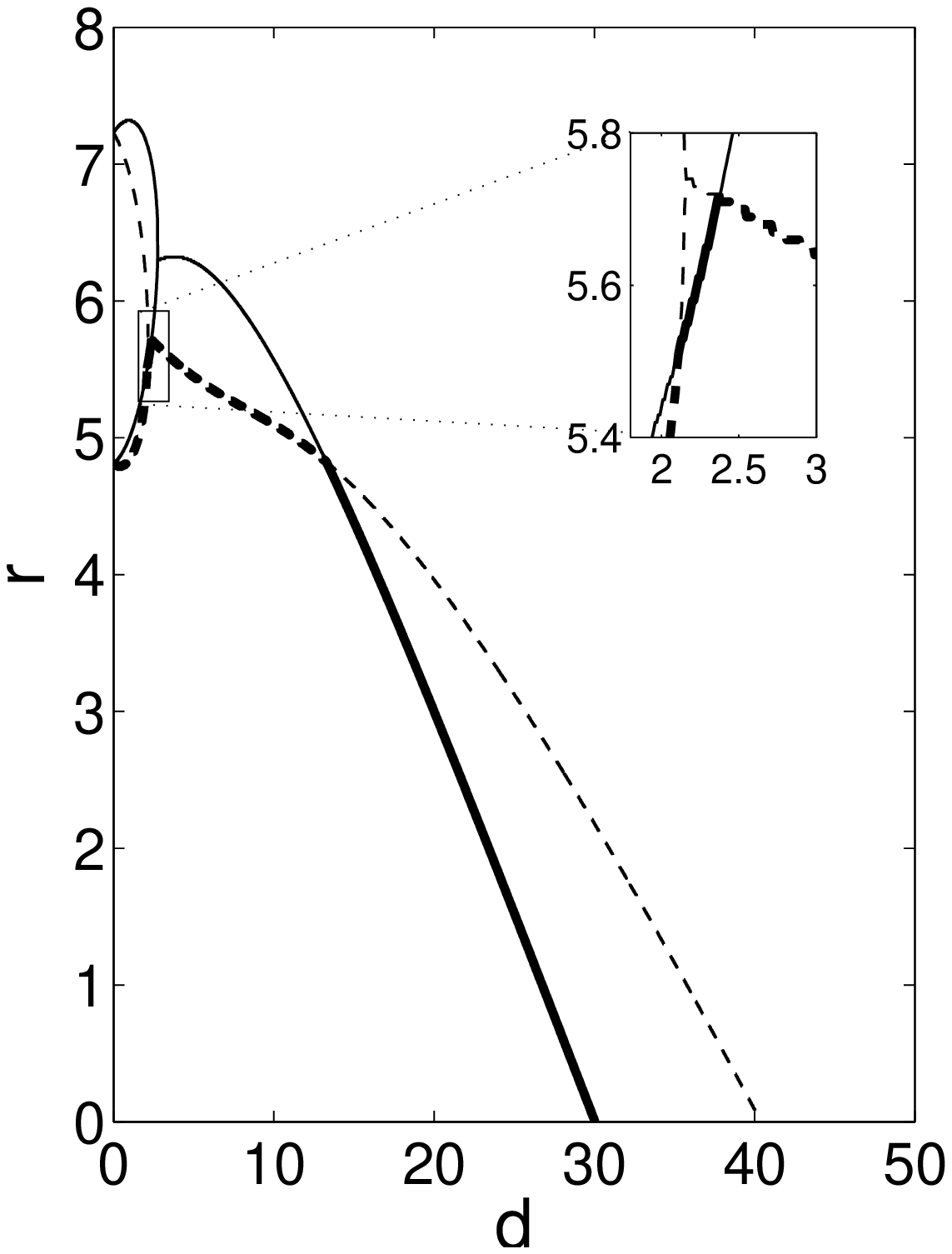}}}\\
\subfigure[$l=10$]{
\resizebox*{5cm}{!}{\includegraphics{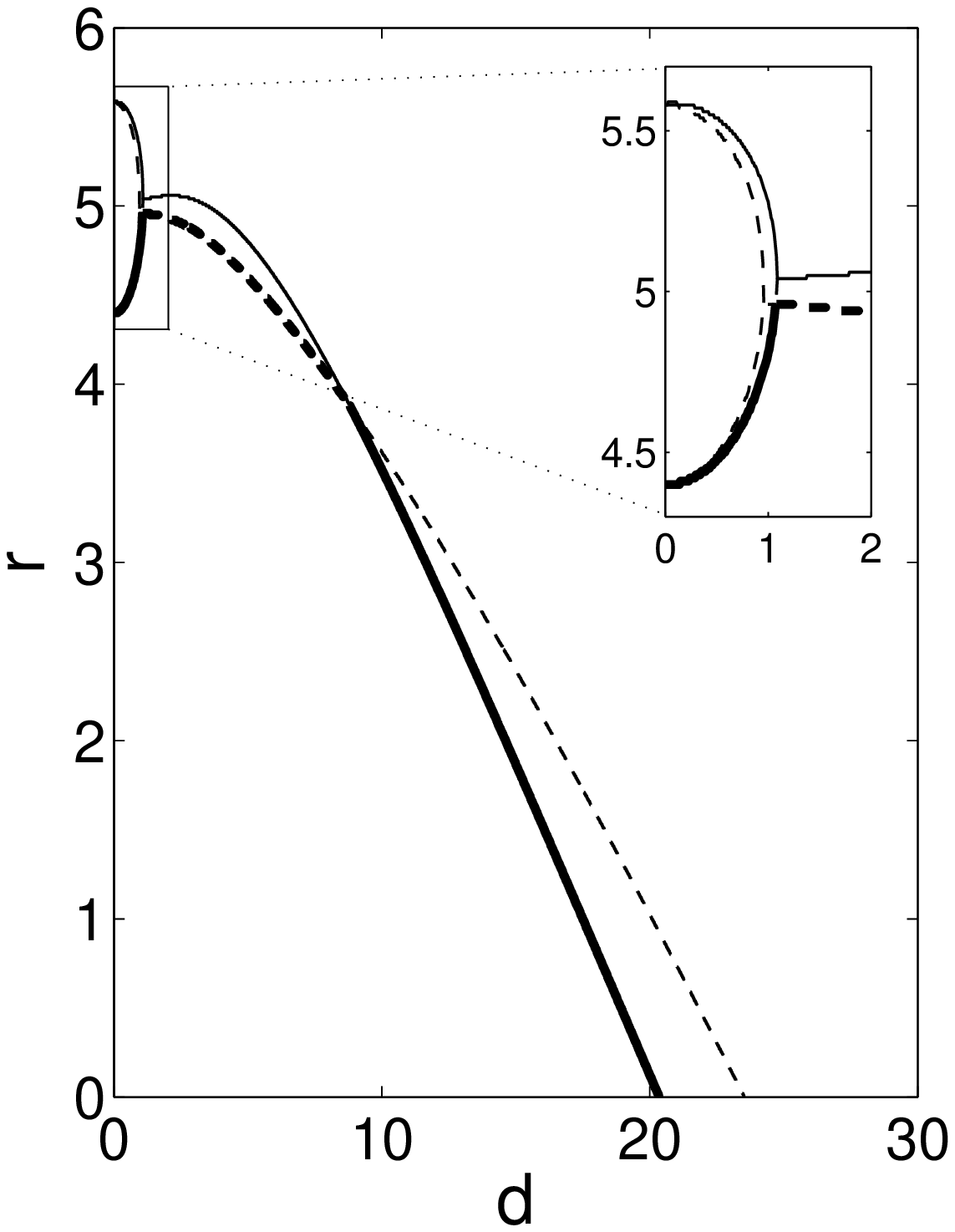}}}%
\subfigure[$l=100$]{
\resizebox*{5cm}{!}{\includegraphics{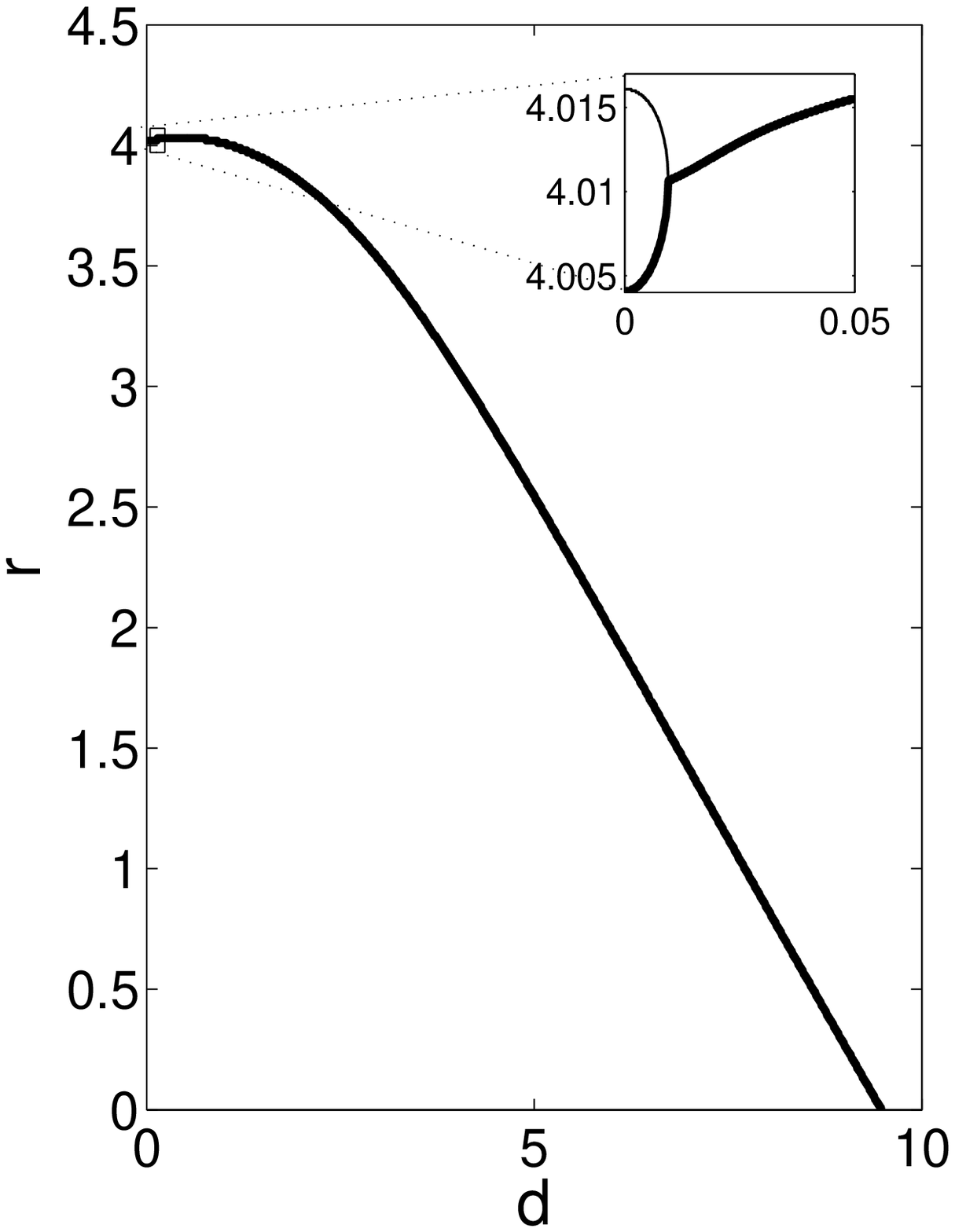}}}%
\caption{Linear stability boundaries for dipole (solid line) and quadrupole (dashed line) parities, for varying values of $l$.  First and second modes of stability are plotted. For sufficiently small $d$ the eigenvalues are real. For each parity the curves for the two modes join at a codimension-2 point, after which the eigenvalues are complex.  For these two lines, the boundary with the lower $r$ value at each $d$ is shown in bold, so that the bold line gives the marginal stability boundary.  The more complex parity changes in figures (b) and (c), and the codimension-2 point at small $d$ in figure (d), are enlarged in their respective figures.} \label{fig:sep_par_bold}
\end{center}
\end{figure}

The marginal curves for the first mode (one zero eigenvalue) and second mode (one positive and one zero eigenvalue) are calculated and plotted for each parity, as functions of $d$ and $r$  for varying values of $l$, in figure \ref{fig:sep_par_bold}.  The marginal curves for the two modes (corresponding initially to a zero eigenvalue) join at a codimension-2 point (double zero eigenvalue), and for larger values of $d$ the eigenvalues on the marginal curve are purely imaginary.  The critical values of $r$ at $d=0$ can be calculated exactly for this simple problem, giving $r=r_n = 4 \left( 1 + n^{2} \pi^{2}/l^{2}\right)$, irrespective of parity, where $n=1$ and $n=2$ correspond to the first and second modes of stability respectively.  We deduce that as $l$ increases, $r_1,r_2 \rightarrow 4$, so that the gap between the two modes decreases, as shown in figure \ref{fig:sep_par_bold}. \\

The bolder lines in figure \ref{fig:sep_par_bold} represent the marginal solution boundary; the least values of $r$ for each $d$ for marginal solutions.  There are changes in parity on this boundary at the points where the dipole and quadrupole branches cross, however for all $l$, and large enough $d$, dipole modes are preferred at onset.  The more complex parity changes are enlarged in the figures.  \\

\begin{figure}
\begin{center}
\includegraphics[width=6cm]{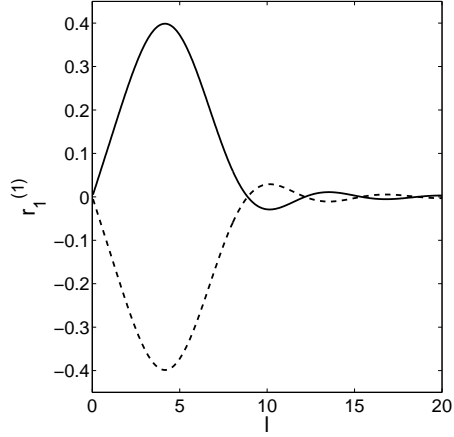}
\caption{The gradient of the marginal curve for the first mode of stability, ${r_{1}}^{(1)} $, at $d=0$, plotted as a function of $l$.  Dipole (solid line) and quadrupole (dashed line) parities have equal and opposite gradients.  Quadrupole parity is preferred at small $l$, and the parity of the marginal curve oscillates for larger $l$.}\label{fig:initial_slope}
\end{center}
\end{figure}

All calculations have been performed for positive values of $d$ as these are the physically most relevant. It is however interesting to note that due to an adjointness property of the governing equations the stability boundaries for negative $d$ are the mirror images of those in figure \ref{fig:sep_par_bold} with dipole and quadrupole modes interchanged \citep{Proctor77}.  It follows that in figure \ref{fig:sep_par_bold} the initial gradients of the marginal curves at small $d$ for the two parities are equal and opposite for all values of $l$.  To calculate the gradient of the stability boundary for the first mode, $r_{1}$, at $d=0$, we expand $A$, $B$ and $r$ in powers of $d$ in equations (\ref{eq:linear_eqn_A}, \ref{eq:linear_eqn_B}).  This results in the formula $r_{1} = {r_{1}}^{(0)} + {r_{1}}^{(1)} d$, where ${r_{1}}^{(0)}$ is the value of $r_{1}$ when $d=0$, and ${r_{1}}^{(1)} $ is the gradient of the stability boundary at $d=0$.  The gradient ${r_{1}}^{(1)} $ is plotted as a function of $l$ in figure \ref{fig:initial_slope} for both parities.  The preferred mode at onset corresponds to the smallest value of $r_{1}$, hence quadrupole modes are preferred at smaller values of $l$, and for larger values of $l$ the initial parity of the marginal curve oscillates. \\

\begin{figure}
\begin{center}
\subfigure[$l=4$]{
\resizebox*{5cm}{!}{\includegraphics{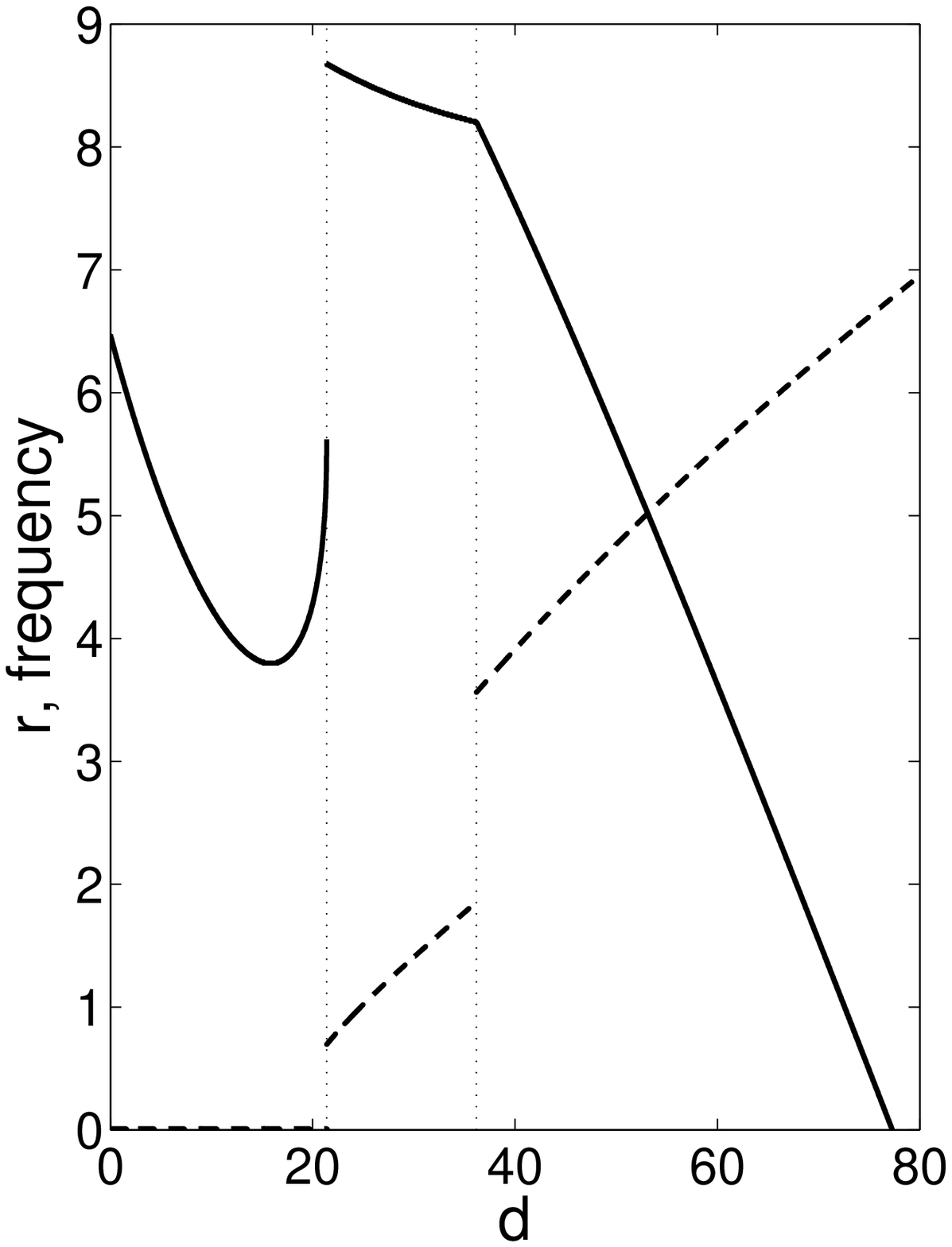}}}%
\subfigure[$l=7$]{
\resizebox*{5cm}{!}{\includegraphics{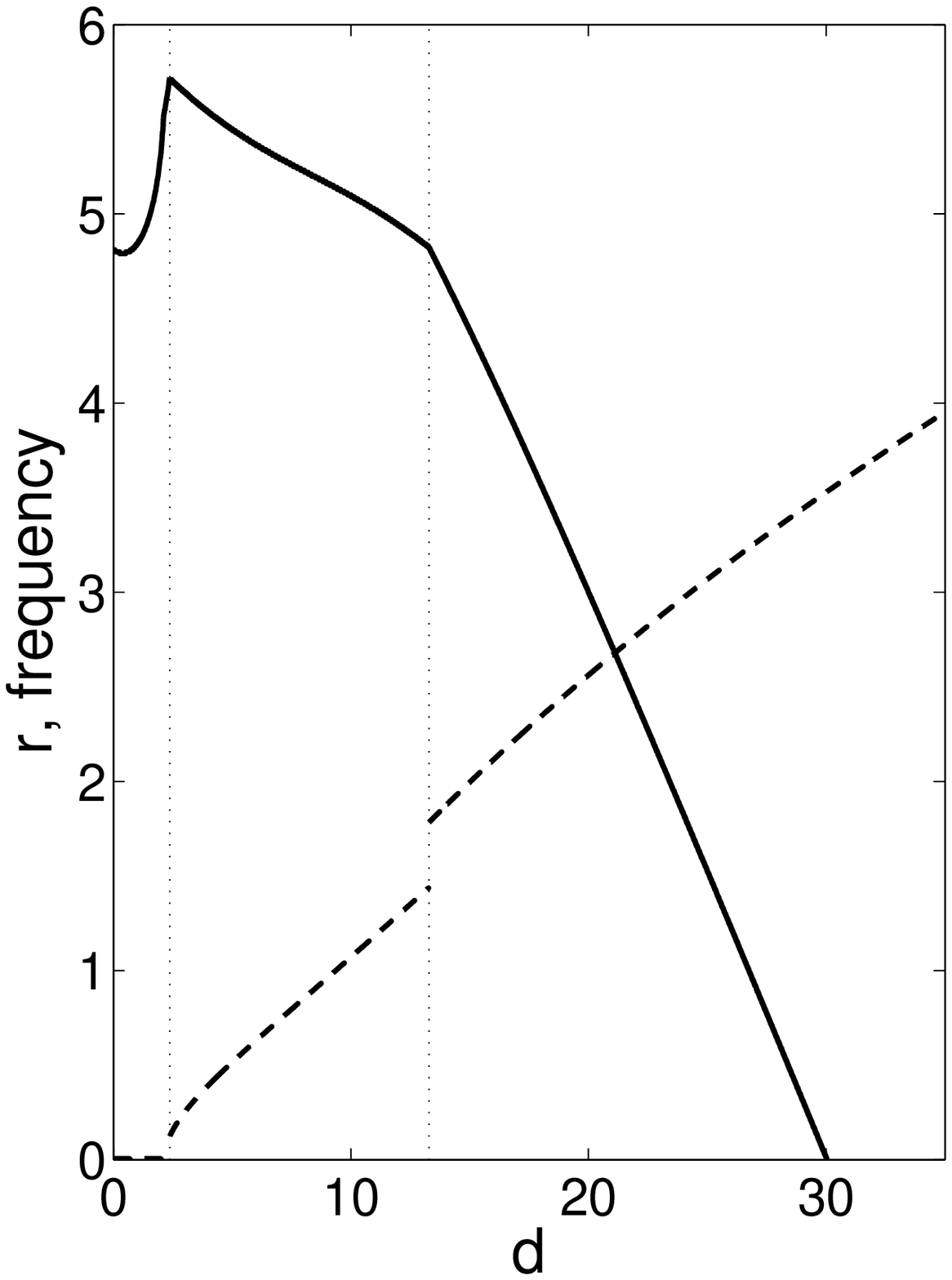}}}%
\subfigure[$l=10$]{
\resizebox*{5cm}{!}{\includegraphics{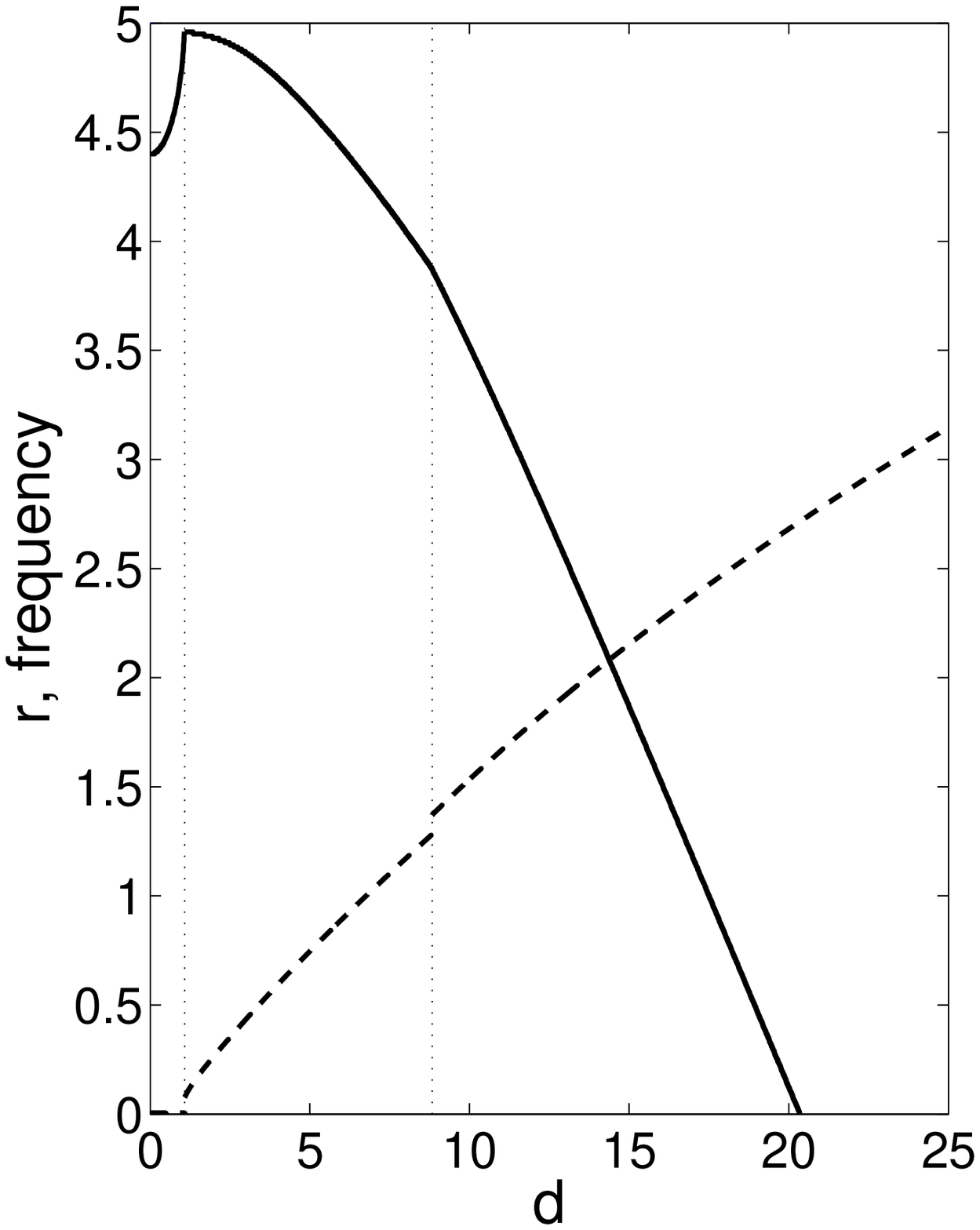}}}%
\caption{The least value of $r$ for unstable solutions for each value of $d$ is plotted (solid line) for varying values of $l$.  These lines are the bold lines from figure \ref{fig:sep_par_bold}.  Also plotted is the corresponding frequency (dashed line) for each point on the line.  Changes in the parity of the marginal solution boundary align with discontinuities in the frequency.  Points on the marginal solution boundary before the codimension-2 point have zero frequency.} \label{fig:stab_and_freq}
\end{center}
\end{figure}

The frequency of the modes on the marginal solution boundary is zero before the codimension-2 point and increases monotonically afterwards.  This corresponds to steady solutions initially and oscillatory solutions after the codimension-2 point.  In figure \ref{fig:stab_and_freq}, the marginal solution boundary from figure \ref{fig:sep_par_bold} is plotted (solid line), along with the corresponding frequency (dashed line).  Vertical lines match up changes in the frequency with changes in parity of the marginal solution boundary. \\

\subsection{Nonlinear Model} \label{sec:nonlinear_model}

To investigate the behaviour of the system in the nonlinear regime, we consider the full nonlinear system given by (\ref{eq:numerical_eqn_A}, \ref{eq:numerical_eqn_B}) which include the simplest form of $\alpha$ quenching for both the mean and fluctuating parts of the $\alpha$-effect.  It transpires later that the direct application of the fluctuating theory to this nonlinear system does in fact lead to a different nonlinearity in the $r$ term, see (\ref{eq:adapted_numerical_A}, \ref{eq:adapted_numerical_B}), however, solutions with this different nonlinearity are qualitatively much the same.  \\

These equations are solved using a second order finite difference scheme, in a box of size $0<x<l$, where $x=0$ represents the North pole, and $x=l$ represents the South pole.  This allows us to define the two parities such that dipole parity corresponds to $B$ odd and $A$ even about the equator ($x=l/2$), and quadrupole parity corresponds to $B$ even and $A$ odd.  $A$ and $B$ are set to be zero at the boundaries, and $d$ and $r$ are varied to investigate stability and parity changes in $d-r$ parameter space.  In all of the nonlinear investigations a value of $l=10$ is used.\\

\begin{figure}
\begin{center}
\subfigure[]{
\resizebox*{5cm}{!}{\includegraphics{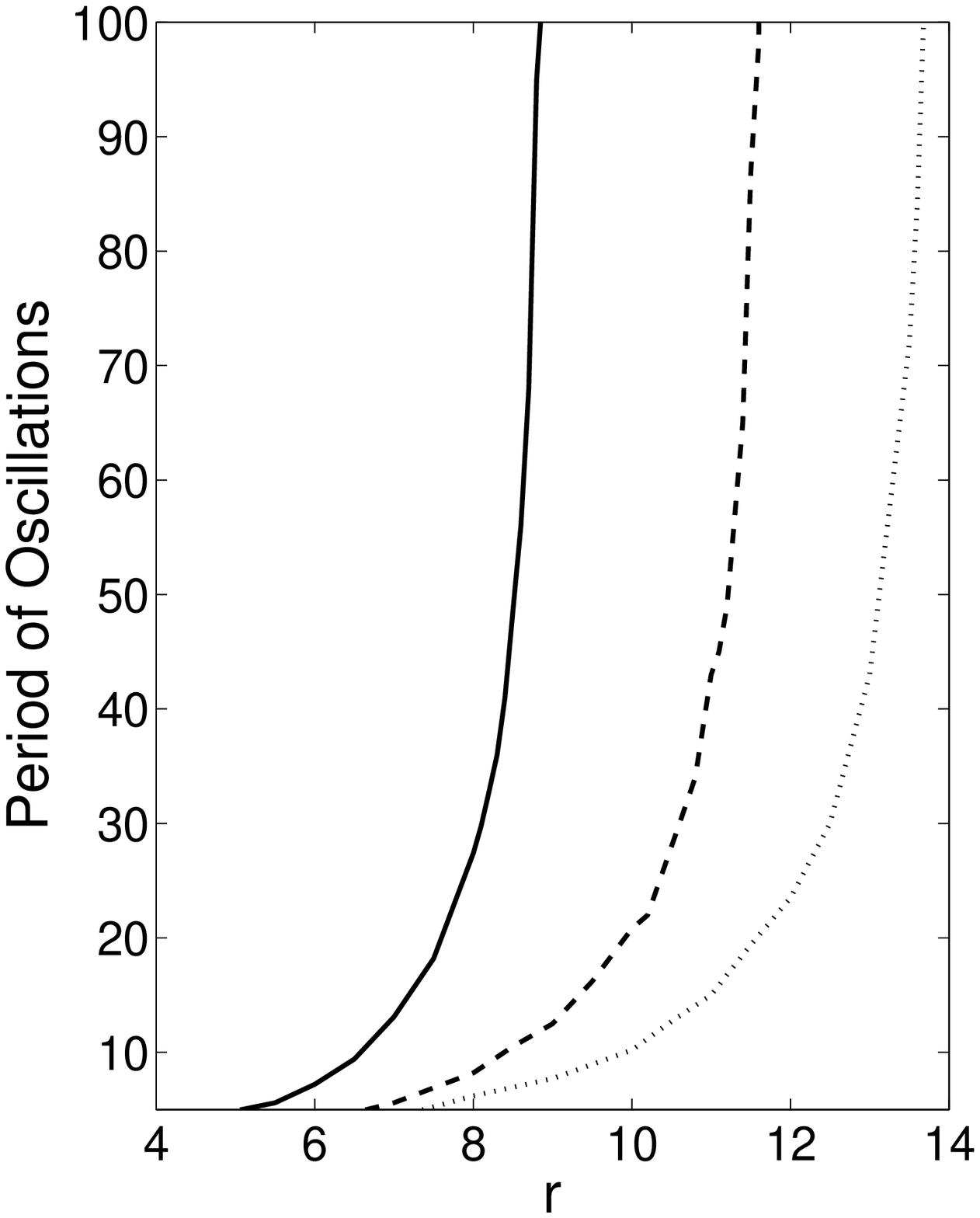}}}
\subfigure[]{
\resizebox*{5cm}{!}{\includegraphics{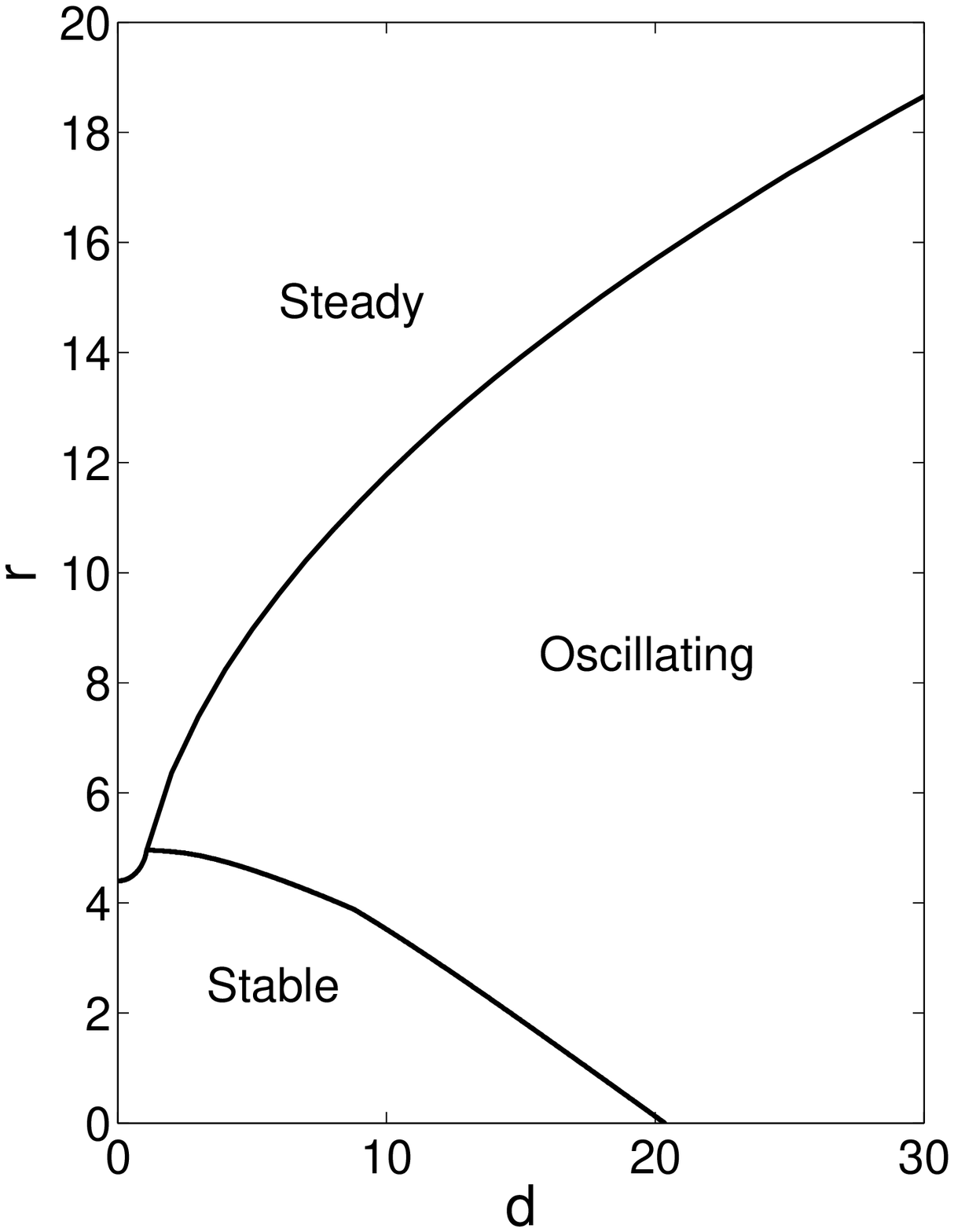}}} 
\caption{(a) Period of oscillations plotted against $r$ (the amplitude of the fluctuating $\alpha$), for varying values of $d$ (the amplitude of the mean $\alpha$).  The values of $d$ are $d=5$ (solid line), $d=10$ (dashed line), and $d=15$ (dotted line). (b) Regions of different stability in $d$-$r$ phase space.} \label{fig:stability_regions}
\end{center} 
\end{figure}

\cite{Proctor07} found that for a fixed value of $d=30$, and varying values of $r$, both oscillating and steady solutions can be found.  In fact, the period of the oscillating solutions increases with increasing values of $r$, and steady solutions are achieved only when $r$ is sufficiently large.  Further investigation reveals that varying the value of $d$, the critical value of $r$ to achieve steady solutions, $r_{c}$, also changes.  Furthermore, the larger the value of $d$, the larger the value of $r_{c}$.  This is shown in figure \ref{fig:stability_regions}(a) where the period of oscillations is plotted as a function of $r$ for three different values of $d$.  In this figure, a steady solution is defined as one with infinite period.  The increase of $r_c$ with $d$ is clear.\\
 
The onset of steady solutions in $d-r$ space is plotted in figure \ref{fig:stability_regions}(b).  The marginal stability boundary for $l=10$ from the linear model has also been included in the figure and the regions of different types of solution are labelled.  As $r$ increases through the oscillating region, the period of the solutions increases until it becomes infinite and the solution reaches a steady state.  \\

\begin{figure}
\begin{center}
\subfigure[]{
\resizebox*{6cm}{!}{\includegraphics{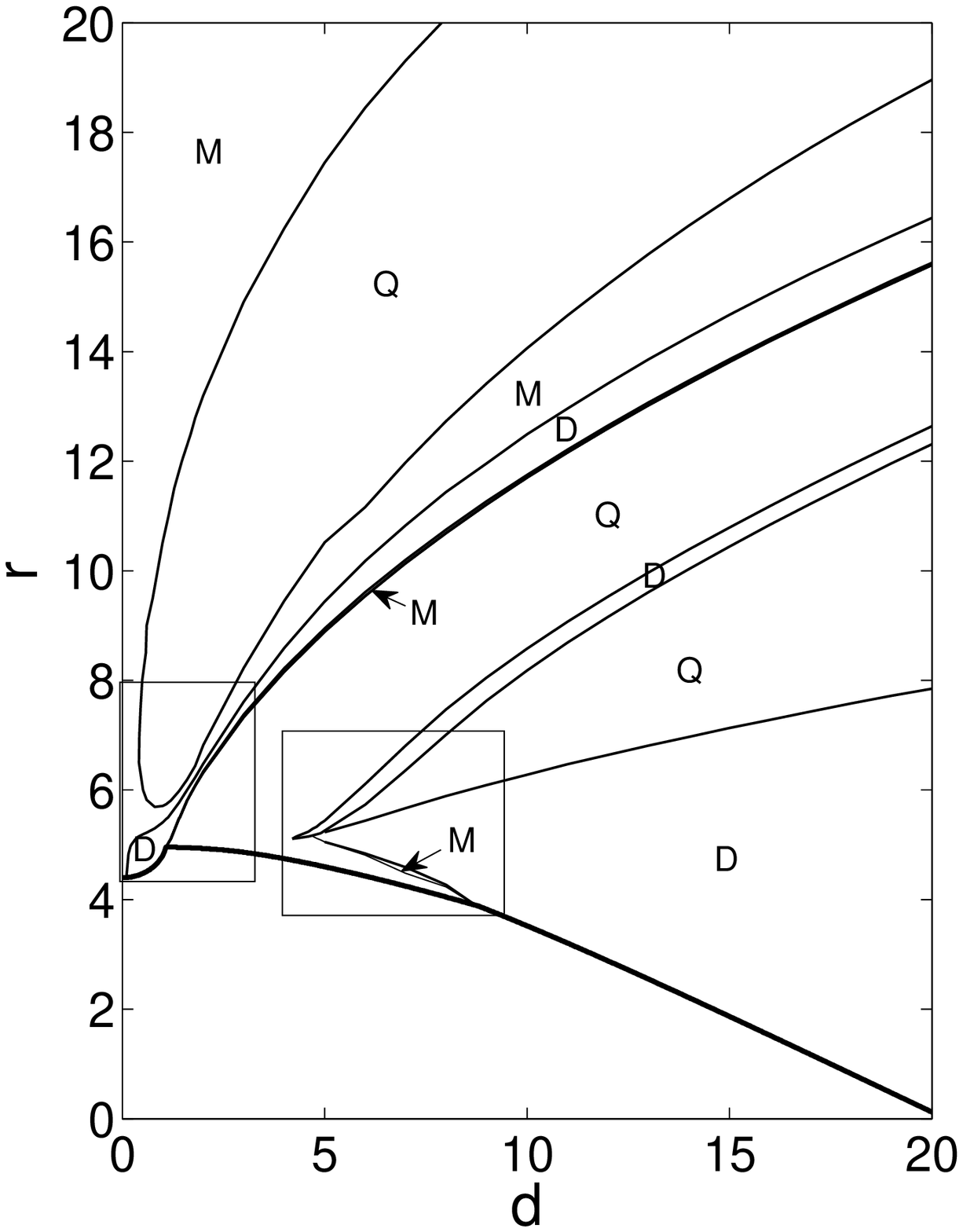}}}\\
\subfigure[]{
\resizebox*{5cm}{!}{\includegraphics{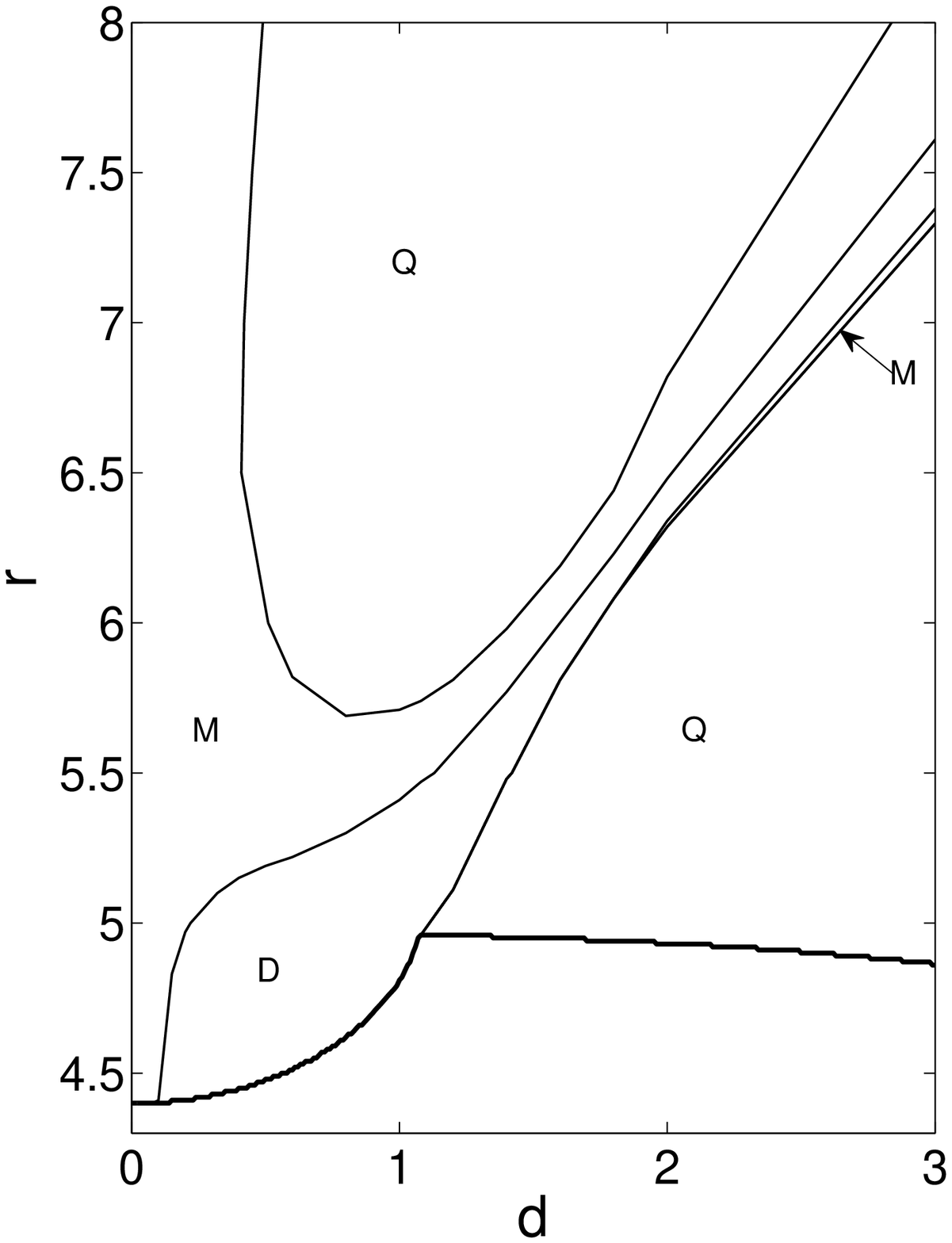}}}%
\subfigure[]{
\resizebox*{6cm}{!}{\includegraphics{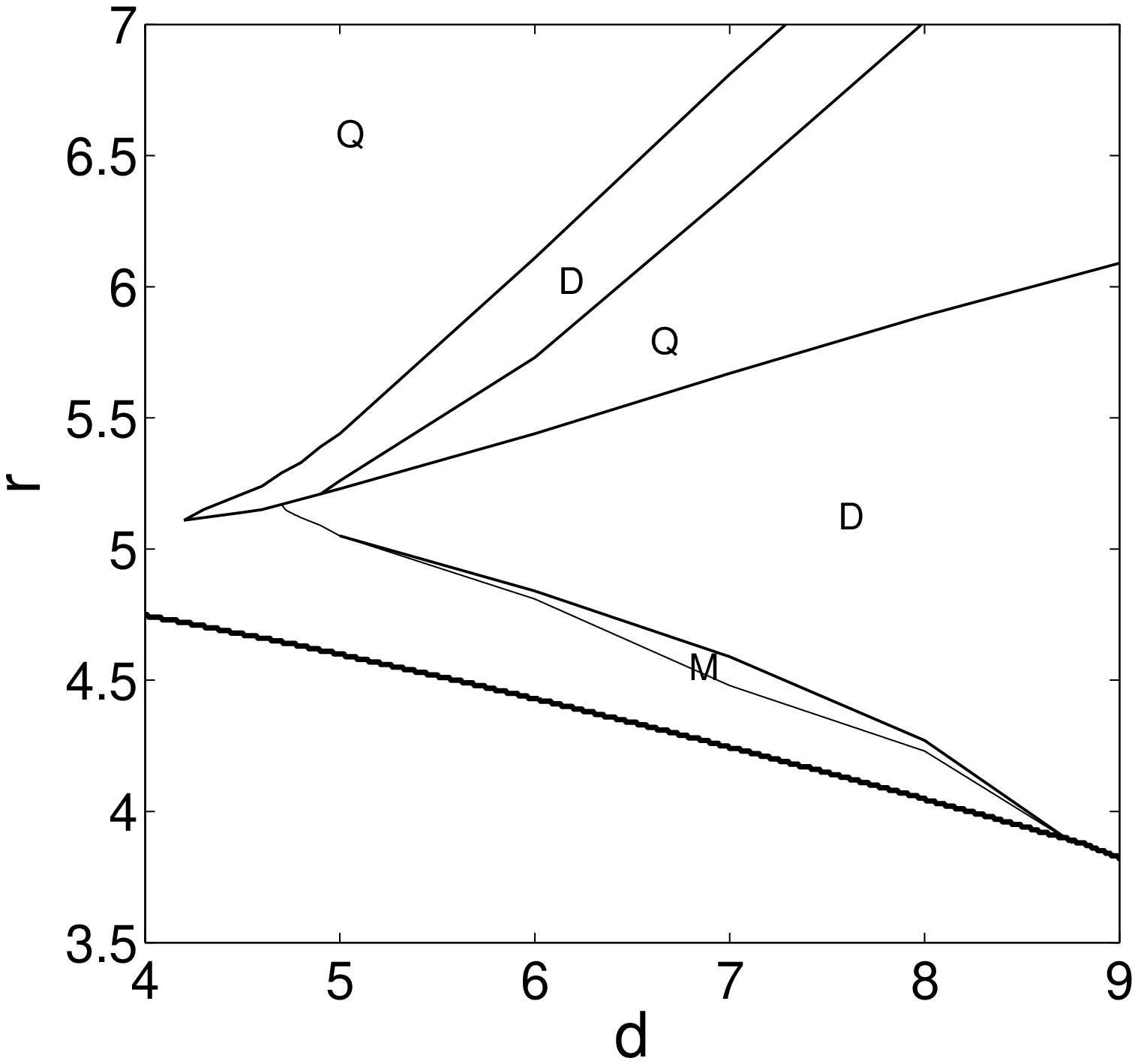}}}%
\caption{(a) Regions of stable parity throughout phase space.  The two boxes in figure (a) are enlarged in figures (b) and (c) below.} \label{fig:parity_regions}
\end{center}
\end{figure}

Mixed initial conditions were used in figure \ref{fig:stability_regions}(b) so that neither $A$ nor $B$ are even or odd about the equator.  This allows solutions to settle into their preferred parity after integrating for long times.  A parameter study reveals that there are many different regions of stable parity throughout parameter space.  These are shown in figure \ref{fig:parity_regions}(a), where the boxes are enlarged in figures \ref{fig:parity_regions}(b) and \ref{fig:parity_regions}(c) to show the more detailed structure.  There are large regions of mixed parity, where the solution does not settle into either dipole or quadrupole parity after integrating for long times, between regions of dipole and quadrupole parity in the steady regime.  However, the only region of mixed solutions in the oscillating region is that in figure \ref{fig:parity_regions}(c) separating the initial parity from the stability boundaries.  We see that this region of mixed solutions meets the stability boundary at exactly the same point as where the two separate parity stability boundaries cross (seen in figure \ref{fig:sep_par_bold}). \\

It is possible to fix the solutions to be either dipole or quadrupole by integrating equations (\ref{eq:numerical_eqn_A}, \ref{eq:numerical_eqn_B}) over half of the domain.  It is then necessary to fix the boundary conditions so that 
\begin{eqnarray*}
A &=& 0 \hspace{0.5cm} \mbox{at} \hspace{0.5cm} x=0, \\
B_{x} &=& 0 \hspace{0.5cm} \mbox{at} \hspace{0.5cm} x=l/2,
\end{eqnarray*}
for a dipole solution, and 
\begin{eqnarray*}
A_{x} &=& 0 \hspace{0.5cm} \mbox{at} \hspace{0.5cm} x=0,\\
B &=& 0 \hspace{0.5cm} \mbox{at} \hspace{0.5cm} x=l/2,
\end{eqnarray*}
for a quadrupole solution.  \\

\begin{figure}
\begin{center}
\includegraphics[width=6cm]{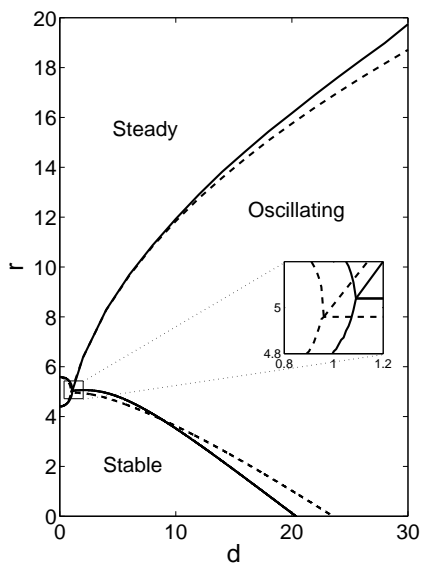}\\
\caption{Regions of different stability for fixed dipole (solid) and quadrupole (dashed) parities.  The onset of steady dynamo action is plotted for each parity and the stability boundaries from figure \ref{fig:sep_par_bold}(c) are added.  The region where the steady boundary meets the stability boundary at the codeimension-2 point for each parity is enlarged.} \label{fig:stability_regions_both_parities}
\end{center}
\end{figure}

Figure \ref{fig:stability_regions_both_parities} shows the onset of steady solutions for the two separate parities, obtained by using these boundary conditions.  The critical value of $r$ to achieve steady solutions is lower for quadrupole parity solutions than it is for dipole parity solutions.  In this figure, the stability boundaries for both dipole and quadrupole parities from figure \ref{fig:sep_par_bold}(c) are added.  For each parity, the line plotting the onset of steady solutions meets the stability boundary at the codimension-2 point, which is enlarged in the figure.  This reinforces the fact that steady solutions are found to the left of this point, and oscillatory solutions are to the right. \\

\section{Testing the asymptotic model with a rapidly varying $\alpha$ function} \label{sec:test_with_random_function}

The theory leading to the model we have studied can be tested against a simple one-dimensional dynamo model (similar to (\ref{eq:1d_model_A}, \ref{eq:1d_model_B}) adapted to be solved numerically in the same way as in section \ref{sec:numerical_model}) where $\alpha$ is now a fluctuating function of time with mean value $d$ and fluctuating part $\~{\alpha}$.  By increasing the magnitude of $\~{\alpha}$, we expect to see the cycle period increase, and if the theory is correct there should be a predictable relation between $\~{\alpha}$, $r$ and the increase in period.\\

The model used to test the theory is
\begin{subequations}
\begin{eqnarray}
A_{t} &=& \frac{(-d \sin (2 \pi x/l)+ \delta^{-2} \~{\alpha}) B}{1 + B^{2}} + A_{xx} - A, \label{eq:random_alpha+d_A}\\ 
B_{t} &=& A_{x} + B_{xx} - B, \label{eq:random_alpha+d_B}
\end{eqnarray}
\end{subequations}
where $d$ is the same mean $\alpha$ term as in the fluctuating theory, $\delta$ is a large parameter to ensure $\~{\alpha}$ is large (the same scaling as is used in section \ref{sec:extension_of_model}), and saturation is included so that these equations can be solved numerically, and to ensure the magnetic fields are quenched.\\

\begin{figure}
\centering
\includegraphics[width=6cm]{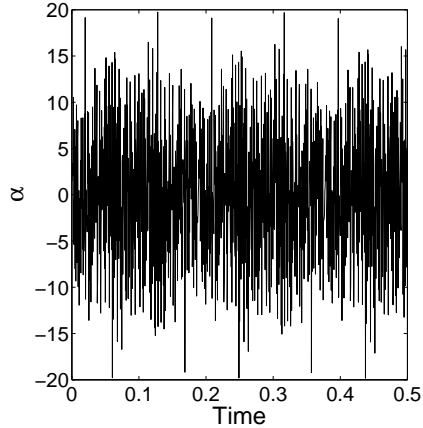}
\caption{Time series for $\~{\alpha}$ with $\zeta=1$, $i_{max}=100$ and $\sigma=100$.} \label{fig:alpha_tilde}
\end{figure}

The form of $\~{\alpha}$ is given by
\begin{equation}
\~{\alpha}=\zeta \sum_{i=1}^{i_{max}} \alpha_{i} \sin (i \sigma t), \label{eq:alpha_tilde}
\end{equation}
where $\sigma$ is the frequency, which is large such that $\sigma = \delta^{-2}\~{\sigma}$ where $\~{\sigma}$ is on order 1 quantity, $\zeta$ is a parameter used to vary the amplitude of the function,  and $\alpha_{i}$ is a sample of randomly generated numbers from a normal distribution with unit standard deviation.  A time series for $\~{\alpha}$ is plotted in figure \ref{fig:alpha_tilde}.\\

To be able to understand the relationship between $\~{\alpha}$ and $r$ in our fluctuating theory, we recall from \cite{Proctor07}, where a time-dependent fluctuating $\alpha$-effect was used, that $\<\alpha_{1} B_{1}\> = -G^{2} \Omega^{\'} B_{x}$, and recalculate $\<\alpha_{1} B_{1}\>$ using equation (\ref{eq:alpha_tilde}) for $\alpha_{1}$.  Recalling that when transforming to the numerical model, $\Omega^{\'} = \eta = 1$ and $G^{2} \rightarrow r$, we find that $r$ and $\zeta$ are related as follows
\begin{equation}
r = \frac{\zeta^{2}}{2} \sum_{i=1}^{i_{max}} \left(\frac{\alpha_{i}}{i\~{\sigma}}\right)^{2}. \label{eq:analytic_r_prediction}
\end{equation}
This expression predicts the corresponding values of $r$ and $\zeta$ that should give solutions with the same period.  Increasing the value of $r$ or $\zeta$ according to this expression should result in the same increase in period.  

\subsection{New form of nonlinear fluctuating $\alpha$ model} \label{sec:new_nonlinear_model}

In order to provide a quantitative comparison between the models in the nonlinear regime it is necessary to derive a new more accurate set of equations for the fluctuating theory.  This is because the new model, (\ref{eq:random_alpha+d_A}, \ref{eq:random_alpha+d_B}), possesses saturation, whereas saturation is only added to the fluctuating theory when adapting it to the numerical model.  To be able to compare these two models we must rederive the fluctuating theory equations from first principles, with a quenched $\alpha$-effect.  The governing equations are then (compare with (\ref{eq:numerical_eqn_A}, \ref{eq:numerical_eqn_B})):
\begin{subequations}
\begin{eqnarray}
A_{t} &=& (\alpha_{0} + \delta^{-2} \alpha_{1} ) f(B) + \eta (A_{xx} - \ell^{2} A), \\
B_{t} &=& \Omega^{\'} A_{x} + \eta (B_{xx} - \ell^{2} B),
\end{eqnarray} 
\end{subequations}
where $\alpha_{0}$ is the mean part of $\alpha$, $\alpha_{1}$ is the fluctuating part of $\alpha$ and is a function of the small timescale $\tau$, and $f(B)=B /(1+B^{2})$ for our model.  \\

Following the same method as in sections \ref{sec:extension_of_model} and \ref{sec:numerical_model}, a new nonlinear model can be derived (see the Appendix), given by
\begin{subequations}
\begin{eqnarray}
A_{t} &=& -\frac{d \sin(2\pi x/l) B}{1+B^{2}} - \frac{r(1-B^{2})^{2}B_{x}}{(1+B^{2})^{4}} + A_{xx} - A, \label{eq:adapted_numerical_A}\\
B_{t} &=& A_{x} + B_{xx} -B. \label{eq:adapted_numerical_B}
\end{eqnarray}
\end{subequations}
The solutions of this model are qualitatively similar to the original nonlinear model.

\subsection{Validating the asymptotic theory} \label{sec:comparing_models}

The test model with a fluctuating $\alpha$ function, (\ref{eq:random_alpha+d_A}, \ref{eq:random_alpha+d_B}), and the new nonlinear version of the fluctuating theory, (\ref{eq:adapted_numerical_A}, \ref{eq:adapted_numerical_B}), are both solved in the same way as described in section \ref{sec:nonlinear_model}.  We set $l=50$, $d=25$ for both models so that when both the new term in the asymptotic model ($r$) and the new fluctuating function ($\~{\alpha}$) are zero, both models will have the same solution with the same period.  We use $\sigma=100$ in the test model for rapid temporal variations in $\~{\alpha}$, and choose $\delta=0.1$ so that $\~{\sigma}=1$.\\

The analytic prediction, equation (\ref{eq:analytic_r_prediction}), is tested by solving the improved asymptotic model (\ref{eq:adapted_numerical_A}, \ref{eq:adapted_numerical_B}) with a chosen value of $r$, and solving the test model (\ref{eq:random_alpha+d_A}, \ref{eq:random_alpha+d_B}) with the corresponding value of $\zeta$, according to equation (\ref{eq:analytic_r_prediction}), and the periods of both solutions are measured and compared.  Results are shown in Table \ref{table:period_data}, where some entries correspond to choosing $\zeta$ first and then calculating the corresponding value of $r$.\\

\begin{table}
\tbl{A comparison of the period for the asymptotic model for selected values of $r$ with the period of the test model with corresponding $\zeta$ (calculated according to equation (\ref{eq:analytic_r_prediction})).}
{\begin{tabular}{cccc} \toprule 
$r$ & Period for asymptotic theory & Period for test model \\
\colrule
0 & 1.53 & 1.53 \\
0.50 & 1.63 & 1.63 \\
2 & 1.86 & 1.88 \\
2.02 & 1.88 & 1.85 \\
4 & 2.18 & 2.17 \\
4.54 & 2.28 & 2.30 \\
6 & 2.52 & 2.42 \\
8 & 2.82 & 2.73 \\
8.07 & 2.84 & 2.73 \\
10 & 3.13 & 3.00 \\
12 & 3.40 & 3.15 \\
\botrule
\end{tabular}}
\label{table:period_data}
\end{table}

\begin{figure}
\begin{center}
\subfigure[]{
\resizebox*{6cm}{!}{\includegraphics{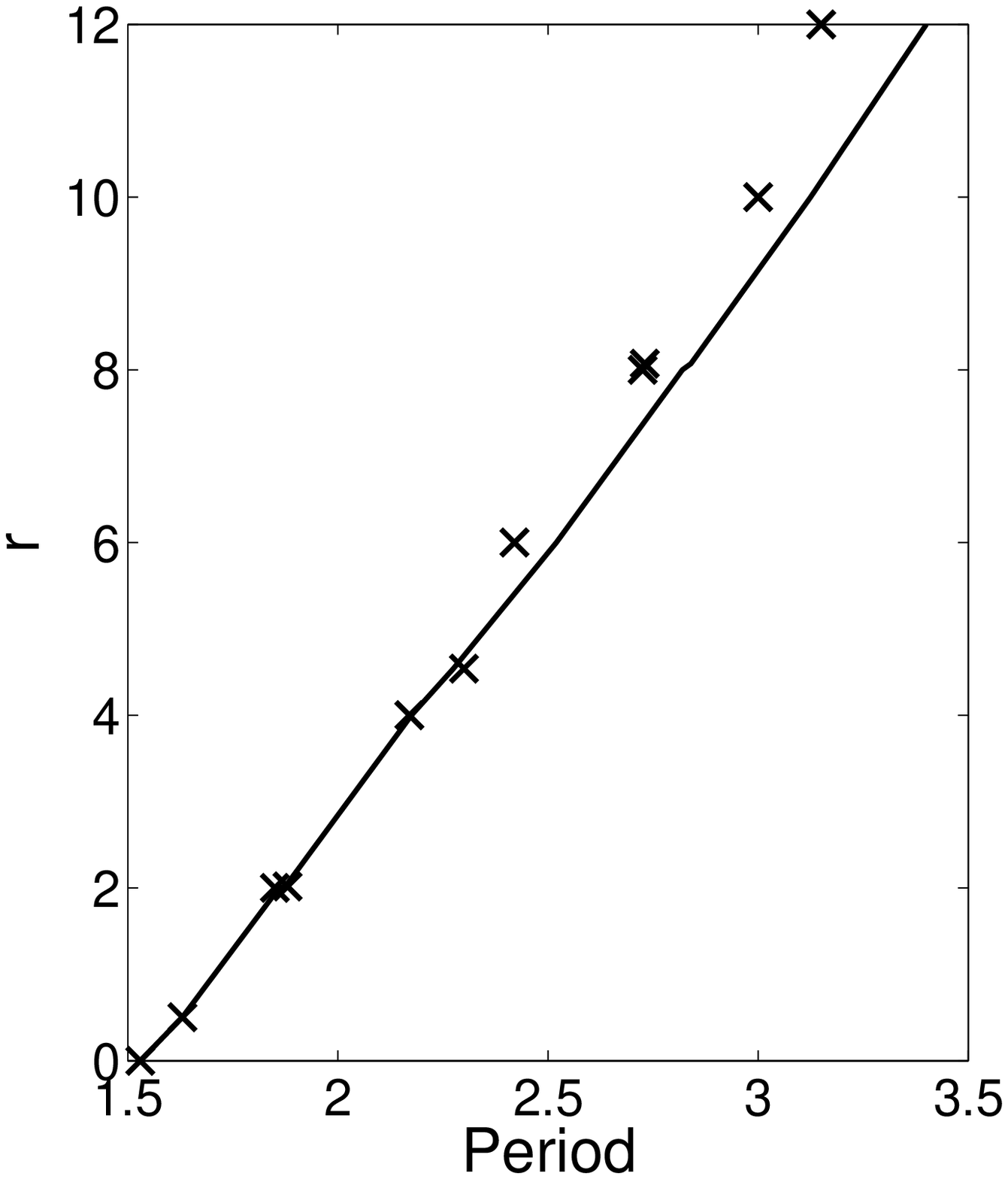}}}
\subfigure[]{
\resizebox*{6cm}{!}{\includegraphics{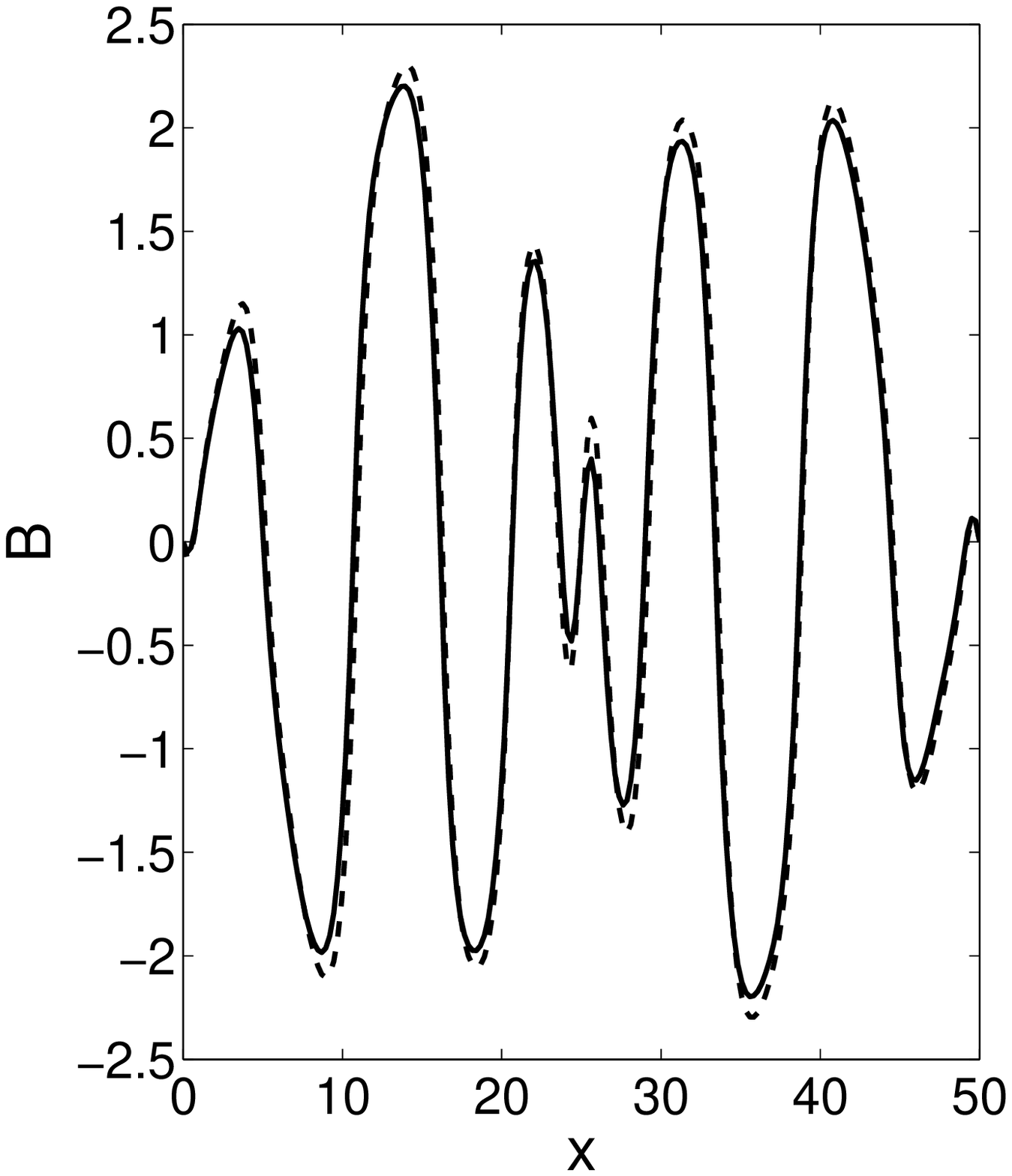}}}
\caption{(a) A comparison of $r$ against period for the improved nonlinear model (solid line) and the test model (crosses), using data from Table \ref{table:period_data}.  The periods are very similar for small values of $r$, but as $r$ increases, the difference between the periods for the two models also increases. (b) A comparison of the toroidal field $B$ plotted against $x$ for the improved nonlinear model (solid line) and the test model (dashed line) for $r = 10$ and $\zeta = 4.45$.} \label{fig:period_comparison}
\end{center}
\end{figure}

The results in Table \ref{table:period_data} are plotted in figure \ref{fig:period_comparison}(a).  The solid line represents the periods from the improved nonlinear model, and the results from the test model are plotted as crosses.  It is clear that when $r$ is small, the models are very similar, however, the difference between the periods gets larger as the solution moves towards a steady state at larger values of $r$.  In figure \ref{fig:period_comparison}(b), for $r=10$, where the difference in period between the two models is sufficiently large, the toroidal field, $B$, is plotted against $x$.  The improved nonlinear model (solid line) and the test model (dashed line) are very much the same, however they do differ slightly, hence the difference in period of these solutions.  \\

A possible explanation for the difference between the models at large $r$ is that the asymptotic theory doesn't allow for the fact that $A$ and $B$ have to vanish at the boundaries, so there are likely to be small errors associated with this.  In an attempt to reduce this error, we have used a larger value of $l$, $l=50$, which has minimised the boundary effects.  However, the general conclusion is that the overall effect of increasing the amplitude of $\~{\alpha}$ in the test model is an increase in the period of the solution, and therefore the test model seems to correctly reproduce the results from the time-dependent asymptotic $\alpha$ theory at small $r$.  

\subsection{Spatiotemporal Case} \label{sec:spatiotemporal_case}

Since one of the aims of this paper is to investigate the addition of spatial dependence in the $\alpha$-effect fluctuations, it is of interest to extend the verification of the theory to the spatiotemporal case.  We now take $\~{\alpha}$ to be
\begin{equation}
\~{\alpha}=\zeta \sum_{i=1}^{i_{max}} \sum_{j=1}^{j_{max}} \alpha_{ij} \sin \left( i \sigma t \right) \sin \left( j \kappa x \right), \label{eq:alpha_tilde_space}
\end{equation}
where $\zeta$, $\alpha_{ij}$ and $\sigma$ are the same as in equation (\ref{eq:alpha_tilde}), and $\kappa = \delta^{-1}\~{\kappa}$ is the spatial fequency, where $\~{\kappa}$ is an order 1 quantity.\\

Unfortunately, a quantitative comparison with the spatiotemporal asymptotic theory is not practical because it is not possible to calculate the contribution to the $\alpha$-effect in a finite geometry (as mentioned in section \ref{sec:extension_of_model}).  However, it is possible to solve equations (\ref{eq:random_alpha+d_A}, \ref{eq:random_alpha+d_B}) with $\~{\alpha}$ as given in equation (\ref{eq:alpha_tilde_space}), in the linear regime (for simplicity), and show that increasing $\zeta$ does indeed increase the growth rate of magnetic energy. \\

\begin{figure}
\centering
\includegraphics[width=6cm]{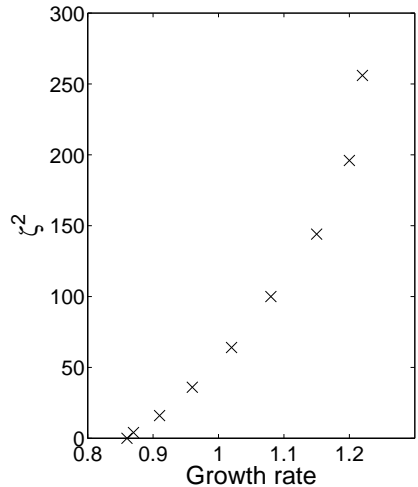}
\caption{$\zeta^{2}$ plotted as a function of the growth rate of magnetic energy for the test model with a spatially and temporally fluctuating $\alpha$-effect.} \label{zeta_vs_growth_rate}
\end{figure}

The theory allows arbitrary additive choices of different modes, but it is simpler to use just one mode for calculations, due to numerical difficulties with resolution, therefore we take $i_{max}=j_{max}=1$.  We set $l=50$, $d=25$ and choose $\~{\sigma} = 1$ and $\~{\kappa}=0.5$; this corresponds to a positive value of $r$, where we would expect to see an increase in the growth rate in the asymptotic theory.  Results are shown in figure \ref{zeta_vs_growth_rate}, where $\zeta^{2}$ is plotted against the growth rate.  Although it is not possible to make a comparison between the two models in the spatiotemporal case, it is clear that an increase in $\zeta$ does lead to an increase in the growth rate of magnetic energy, as predicted by the asymptotic theory.\\

\section{Conclusion} \label{sec:conclusion}

The interaction of a spatially and temporally fluctuating $\alpha$-effect with large-scale shear can lead to a large-scale dynamo mechanism, as in \cite{Proctor07}.  Moreover, a similar term to the new term found in \cite{Proctor07} relating to the fluctuating $\alpha$-effect is calculated and found to be positive or negative, depending on the spatiotemporal spectrum of the fluctuations.  If the term is negative the critical value of the dynamo number is increased, and dynamo action is inhibited.\\

Numerical investigations were made, both in the linear and nonlinear regime, by adapting the equations to a simple one-dimensional model of the solar cycle, and investigating behaviour in the two-dimensional $d-r$ parameter space.  In the linear model it was found that the values of $r$ for the onset of unstable solutions are different for dipole and quadrupole parity systems at a fixed value of $d$, but these become more and more alike as $l$ increases.  The first two modes of stability are initially steady state, but join together at a codimension-2 point, where there is a region of solutions with zero frequency, i.e. steady solutions, to the left of this point, and oscillating solutions to the right. \\

In the nonlinear model, the onset of steady solutions was plotted in $d-r$ space and it was found that an increase in the fluctuating $\alpha$ parameter, $r$, led to an increase in the cycle period of the oscillations.  Regions of stable parity were determined throughout phase space and were found to be quite complicated considering the simplicity of the model.  The boundary conditions were adjusted to allow the determination of the onset of steady and oscillatory solutions for the dipole and quadrupole cases separately, and combining these with the corresponding stability boundaries from the linear model reveals that the curve separating steady solutions from oscillatory solutions in the nonlinear regime tends to the codimension-2 point where the two types of linear stability join. \\

Finally, the asymptotic theory is verified against a one-dimensional dynamo wave model with a fluctuating function for the $\alpha$-effect.  Firstly, the time-dependent case is verified in the nonlinear regime, where a new more accurate version of the asymptotic theory from \cite{Proctor07} is derived and used to compare with a temporally fluctuating function for $\alpha$.  The relationship between the time-dependent function and the previous $r$ term is found and tested, and we see an increase in the cycle period of the oscillations in the predicted way according to the relationship derived.  Secondly, an attempt to verify the spatiotemporal case is made using a space and time-dependent $\alpha$ function.  It is not possible to make a quantitative comparison with the asymptotic theory in this case, due to contributions to the $\alpha$-effect in the analytic theory which cannot be calculated in a finite domain.  Instead, the test model is solved, in the linear regime, and it is shown that increasing the value of the spatiotemporal $\alpha$ function does indeed increase the growth rate of magnetic energy. \\ 

\appendices 

\section{\vspace{12pt}\\Derivation of improved nonlinear model}

The governing equations are
\begin{eqnarray*}
A_{t} &=& \alpha f(B) + \eta (A_{xx} - \ell^{2} A), \\
B_{t} &=& \Omega^{\'} A_{x} + \eta (B_{xx} - \ell^{2} B).
\end{eqnarray*} 
We follow the derivation of the fluctuating $\alpha$ model used in \cite{Proctor07}, since this is the temporal case, with a more general function of $B$.  We split $\alpha$, $A$ and $B$ into their mean and fluctuating parts and apply a method of multiple scales such that
\begin{eqnarray*}
\alpha &=& \alpha_{0} + \ep^{-1} \alpha_{1}(\tau), \\
A &=& A_{0} + A_{1}(\tau), \\ 
B &=& B_{0} + \ep B_{1}(\tau), \\
f(B) &=& f(B_{0}) + \ep B_{1} f^{\'} (B_{0}) + \cdots ,
\end{eqnarray*}
where subscripts $0$ and $1$ denote the mean and fluctuating parts respectively, $\tau$ is the intermediate timescale and $\ep$ is taken to be small.  A suitable average is taken such that $\<\alpha_{1}\> = \<A_{1}\> = \<B_{1}\> =0$ and $\<f(B)\> = f(B_{0}) + O(\ep^{2})$, and we obtain the following leading order mean field equations
\begin{subequations}
\begin{eqnarray}
{A_{0}}_{t} &=& \alpha_{0} f(B_{0}) + \< \alpha_{1} B_{1} \> f^{\'}(B_{0}) + \eta({A_{0}}_{xx} - \ell^{2} A_{0}), \label{eq:derivation_mean_A}\\
{B_{0}}_{t} &=& \Omega^{\'} {A_{0}}_{x} + \eta({B_{0}}_{xx} - \ell^{2} B_{0}). \label{eq:derivation_mean_B}
\end{eqnarray}
\end{subequations}
The new term $\<\alpha_{1} B_{1} \>$ is calculated in the usual way and found to be
\begin{equation*}
\<\alpha_{1} B_{1} \> = - G^{2} f^{\'}(B_{0}) \Omega^{\'} {B_{0}}_{x}.
\end{equation*}
Substituting $\<\alpha_{1} B_{1} \>$ in Equation (\ref{eq:derivation_mean_A}) and, to use the same notation as section \ref{sec:nonlinear_model} for comparison, we set $\alpha_{0} = -d \sin (2 \pi x / l)$, $G^{2}=r$, $\Omega^{\'} = \eta = \ell = 1$ and drop the zero subscripts.  Equations (\ref{eq:derivation_mean_A}, \ref{eq:derivation_mean_B}) then become
\begin{subequations}
\begin{eqnarray}
A_{t} &=& - d \sin \left(2 \pi x / l \right) f(B) - r (f^{\'}(B_{0}))^{2} B_{x} + A_{xx} - A, \label{eq:derivation_A}\\
B_{t} &=& A_{x} + B_{xx} - B. \label{eq:derivation_B}
\end{eqnarray}
\end{subequations}

In section \ref{sec:new_nonlinear_model}, $f(B) = B/(1+B^{2})$, so we find
\begin{equation*}
f^{\'}(B) = \frac{1-B^{2}}{(1+B^{2})^{2}},
\end{equation*}
and equations (\ref{eq:derivation_A}, \ref{eq:derivation_B}) become
\begin{subequations}
\begin{eqnarray}
A_{t} &=& -\frac{d \sin \left(2 \pi x / l \right) B}{1+B^{2}}  - \frac{r(1-B^{2})^{2}B_{x}}{(1+B^{2})^{4}} + A_{xx} - A, \label{eq:derivation_final_A}\\
B_{t} &=& A_{x} + B_{xx} - B. \label{eq:derivation_final_B}
\end{eqnarray}
\end{subequations}

This derivation only concerns a temporally fluctuating $\alpha$-effect.  It would be impracticable to calculate the corresponding improved nonlinear model in the spatiotemporal case due to the multiple levels of expansion.

\end{document}